\documentclass[11pt]{article}

\usepackage[utf8]{inputenc}
\usepackage[T1]{fontenc}
\usepackage{microtype}
\usepackage{amsmath}
\usepackage{graphicx}
\usepackage[margin=1.8cm]{geometry}
\usepackage[numbers,sort&compress]{natbib}
\usepackage{booktabs}
\usepackage{caption}
\usepackage{titlesec}
\usepackage{authblk}
\usepackage{abstract}
\usepackage{hyperref}

\titleformat{\section}{\normalfont\bfseries}{\thesection.}{0.5em}{}
\titleformat{\subsection}{\normalfont\bfseries\itshape}{\thesubsection}{0.5em}{}
\title{\textbf{Reconstruction Interval Z-Phase Dependence of AI Detection
Sensitivity in CT Lung Nodule Screening}}

\author[1]{Dan Soliman, MS}
\affil[1]{GammaMetric Medical Physics, Independent Medical Physics Consultancy\\
\texttt{dgsoliman@gmail.com}}
\date{}

\begin{document}
\maketitle

\begin{abstract}
\noindent\textbf{Background:} Sensitivity of AI-assisted lung nodule detection
systems is known to vary with CT acquisition parameters including radiation dose,
reconstruction kernel, and slice thickness. However, the dependence of detection
probability on nodule position within the reconstruction cycle --- the z-phase ---
has not, to the author's knowledge, been characterized for deep learning-based
detection systems.

\noindent\textbf{Methods:} A retrospective analysis was performed using the
LIDC-IDRI dataset. Detection results from a previously validated 154-case
perturbation study were re-analyzed to quantify z-phase sensitivity effects. For
each consensus nodule ($\geq$4-reader agreement), z-phase was defined as the
fractional position of the nodule center within the reconstruction cycle, folded
to [0,\,0.5], where 0 indicates a nodule centered on a reconstruction plane and
0.5 indicates a nodule straddling two adjacent planes. Detection sensitivity was
stratified by z-phase bin and reconstruction interval (1\,mm baseline, 3\,mm,
5\,mm), and by the ratio of reconstruction interval to nodule diameter (d/D).

\noindent\textbf{Results:} At 5\,mm reconstruction interval, overall sensitivity
was 71.6\% compared to 84.8\% at 1\,mm baseline. Within the 5\,mm condition,
sensitivity varied by 17.6 percentage points across z-phase bins (61.9\% at phase
0.45 versus 79.5\% at phase 0.25). At 3\,mm reconstruction interval, the z-phase
effect was negligible (4.7\,pp range, 79.3--85.3\%). When pooled across all
conditions and stratified by d/D ratio, sensitivity was 92.4\% for d/D $<$ 0.5,
78.0\% for $0.5 \leq$ d/D $<$ 1.0, and 61.4\% for d/D $\geq$ 1.0, with a
systematic z-phase effect present only in the d/D $\geq$ 1.0 stratum.

\noindent\textbf{Conclusions:} AI detection sensitivity depends not only on
reconstruction interval size but on the ratio of reconstruction interval to nodule
diameter. When this ratio approaches or exceeds 1.0 (as occurs for 3--6\,mm
nodules at 5\,mm reconstruction) z-phase becomes the dominant source of
per-study detection variance. This effectively stochastic effect is variable for
each patient, invisible to protocol-level quality metrics, and not reflected in AI
confidence scores.
\end{abstract}


\section{Introduction}

The effect of CT reconstruction parameters on pulmonary nodule detection has been
studied extensively in the context of human readers and classical computer-aided
detection (CAD) systems. Fischbach et~al.\ demonstrated improved nodule detection
rates with reduced slice thickness in multislice CT \cite{fischbach2003}, Kim
et~al.\ showed that both section thickness and reconstruction interval
independently affect automated CAD sensitivity \cite{kim2005}, and Marten et~al.\
characterized CAD diagnostic performance across variable reconstruction slice
thickness settings \cite{marten2005}. Recently, attention has turned to the
robustness of deep learning-based detection systems under varying acquisition
conditions. Blazis et~al.\ found that reconstruction kernel and slice thickness
affected deep learning CAD recall and precision \cite{blazis2021}, and a
systematic review by Khan et~al.\ identified CT acquisition variability as the
primary driver of AI performance degradation, with 10--20\% AUC reductions
attributable to kernel, dose, and slice thickness variation \cite{khan2026}.
Soliman et~al.\ further characterized sensitivity loss across dose reduction, soft
kernel, and thick-slice conditions using an empirical 154-case perturbation
framework \cite{soliman2026}.

However, to the author's knowledge, no prior work has characterized the
intra-interval, position-dependent stochastic effect (the dependence of AI
detection probability on where a nodule falls within the reconstruction cycle)
specifically for AI-assisted lung nodule detection. Kim et~al.\ reported that
shorter reconstruction intervals improve CAD sensitivity overall \cite{kim2005},
and Mao et~al.\ found that reconstruction kernel affected AI nodule classification
but not detection \cite{mao2025}; however, sensitivity variation as a function of
nodule z-position within a fixed interval has not been quantified.

This distinction matters clinically. A nodule whose center occupies two adjacent
reconstruction planes is represented in each plane by only a partial cross-section,
reducing the volumetric signal available to a 3D detection network. The position of
any given nodule relative to the reconstruction grid is determined by the scanner's
reconstruction offset and the nodule's anatomic location --- both 
unpredictable in clinical deployment. Two patients scanned with identical protocols
--- or the same patient scanned twice --- may therefore have meaningfully different
AI detection probabilities for the same nodule size, with no indication of this
difference in the AI confidence score or in any protocol-level quality flag.

The present study quantifies this z-phase sensitivity effect empirically using the
LIDC-IDRI dataset and a previously validated deep learning detection pipeline, and
introduces the interval-to-diameter ratio as the governing geometric parameter.

\section{Methods}

\subsection{Dataset}

The Lung Image Database Consortium and Image Database Resource Initiative
(LIDC-IDRI) dataset was used \cite{armato2011}. Detection results were drawn from
a previously reported 154-case perturbation study \cite{soliman2026} in which a
MONAI RetinaNet model trained on LUNA16 was evaluated under five acquisition
conditions: baseline (1\,mm reconstruction), 25\% dose reduction, 50\% dose
reduction, soft reconstruction kernel, 3\,mm slice thickness, and 5\,mm slice
thickness. The present analysis focuses on baseline, 3\,mm, and 5\,mm conditions.

\subsection{Nodule Consensus}

Nodule annotations were drawn from LIDC-IDRI XML reading session files. Consensus
nodules were defined as annotation clusters where all four independent radiologist
readers agreed ($\geq$4-reader consensus), using a 15\,mm spatial clustering
radius. This stricter threshold --- compared to the $\geq$3-reader criterion used
in the prior perturbation study \cite{soliman2026} --- selects for the most
unambiguous nodule findings, and explains the higher baseline sensitivity observed
here (84.8\%) relative to the prior report (78.2\%). A confidence threshold of 0.5
was applied to model detections, and a 15\,mm center-to-detection match radius was
used to score each nodule as detected or missed.

\subsection{Z-Phase Definition}

For each consensus nodule, z-phase $\varphi$ was defined as:
\begin{equation}
\varphi = \min\!\left(
  \frac{|z_{\text{nodule}} - z_{\text{origin}}| \bmod d}{d},\;
  1 - \frac{|z_{\text{nodule}} - z_{\text{origin}}| \bmod d}{d}
\right)
\label{eq:zphase}
\end{equation}
where $z_\text{nodule}$ is the nodule center position in mm (derived from
LIDC-IDRI XML annotations), $z_\text{origin}$ is the inferior z-coordinate of the
CT volume (derived from DICOM \texttt{ImagePositionPatient} headers), and $d$ is
the reconstruction interval in mm. The folding operation maps phase values
symmetrically to [0,\,0.5], where $\varphi = 0$ indicates a nodule centered
exactly on a reconstruction plane and $\varphi = 0.5$ indicates a nodule
equidistant between two adjacent planes. Z-phase values were binned into five equal
intervals spanning [0,\,0.5].

\subsection{Nodule Diameter and Interval/Diameter Ratio}

Nodule diameter was estimated from LIDC-IDRI XML annotation data. For each
annotation, the bounding box extent of edgeMap coordinates was computed in x and y
(converted to mm using DICOM pixel spacing), and the z-extent was computed from the
range of imageZposition values across ROIs. Estimated diameter was taken as the
maximum of the xy and z extents, averaged across readers within each consensus
cluster. The interval-to-diameter ratio d/D was defined as the reconstruction
interval divided by the estimated nodule diameter. Nodules with estimated diameter
$\leq$ 0 were excluded from ratio analyses.

\subsection{Statistical Analysis}

Detection sensitivity was computed as the proportion of consensus nodules detected
above the confidence threshold within each z-phase bin and reconstruction interval
condition. Bootstrap confidence intervals (95\%, 2000 resamples) were computed per
bin. For the ratio analysis, all three reconstruction interval conditions were
pooled and stratified by d/D into three bins: d/D $<$ 0.5 (well sampled),
$0.5 \leq$ d/D $<$ 1.0 (critical range), and d/D $\geq$ 1.0 (undersampled).

\section{Results}

\subsection{Overall Sensitivity by Reconstruction Interval}

A total of 408 consensus nodule instances were analyzed across 154 cases. Overall
detection sensitivity was 84.8\% (346/408) at 1\,mm baseline, 82.4\% (336/408) at
3\,mm reconstruction interval, and 71.6\% (292/408) at 5\,mm reconstruction
interval. The 5\,mm condition represented a 13.2 percentage point absolute
reduction from baseline.

\subsection{Z-Phase Sensitivity Dependence}

At 1\,mm baseline, z-phase sensitivity varied from 78.5\% to 98.5\% across bins
with no consistent directional pattern and heavily overlapping 95\% bootstrap
confidence intervals, consistent with the expectation that sub-millimeter
reconstruction intervals do not produce clinically meaningful partial volume effects
for nodules $\geq$3\,mm.

At 3\,mm reconstruction interval, the z-phase effect was negligible: sensitivity
ranged from 79.3\% to 85.3\% (4.7\,pp range) with no systematic trend and
overlapping confidence intervals across all bins.

\begin{figure}[ht]
\centering
\includegraphics[width=\textwidth]{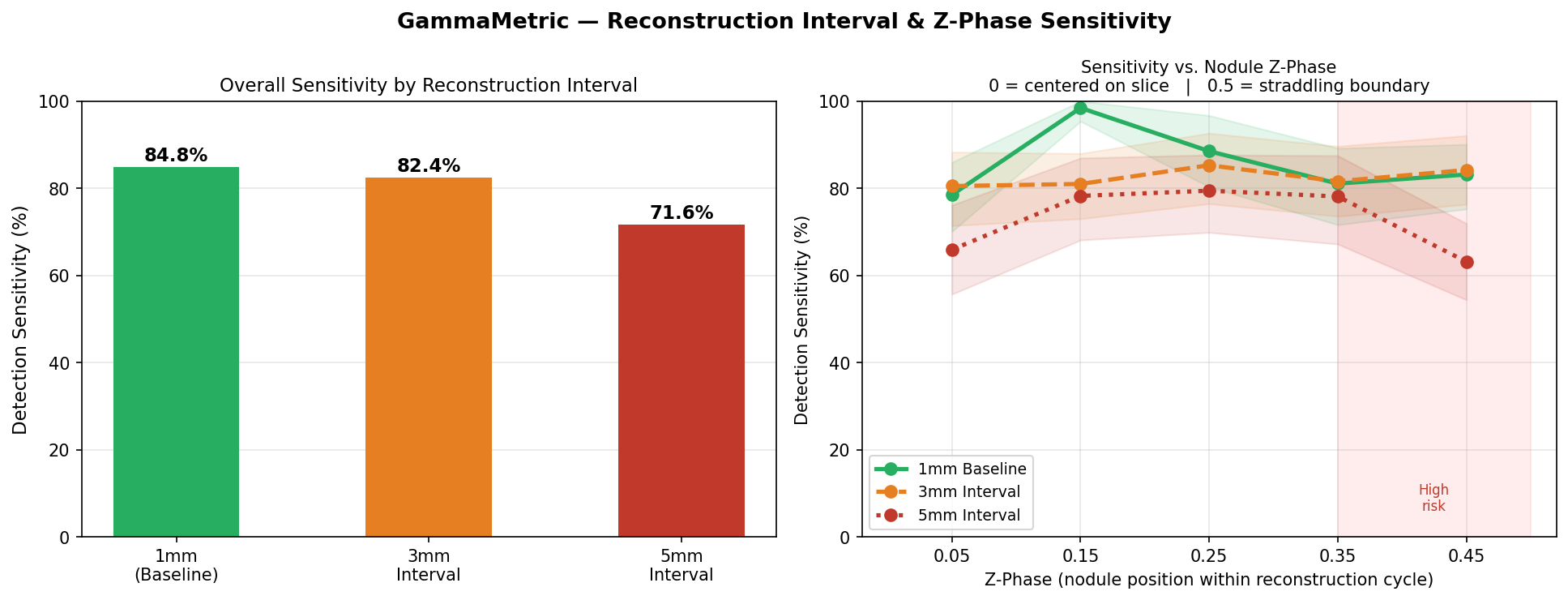}
\caption{\textbf{Reconstruction interval z-phase sensitivity analysis.}
\textit{Left:} Overall detection sensitivity by reconstruction interval (1\,mm
baseline, 3\,mm, 5\,mm) across 408 consensus nodule instances from 154 LIDC-IDRI
cases. \textit{Right:} Detection sensitivity as a function of nodule z-phase within
the reconstruction cycle, stratified by interval, with 95\% bootstrap confidence
intervals shaded. Z-phase $\varphi = 0$ indicates a nodule centered on a
reconstruction plane; $\varphi = 0.5$ indicates a nodule equidistant between
adjacent planes. Shaded region ($\varphi > 0.35$) denotes the high-risk zone.
Detection model: MONAI RetinaNet trained on LUNA16. Consensus criterion:
$\geq$4-reader agreement, 15\,mm clustering radius, 0.5 confidence threshold.}
\label{fig:zphase}
\end{figure}

At 5\,mm reconstruction interval, a clinically meaningful z-phase effect emerged.
Sensitivity was lowest at the extremes of the phase range: 65.9\% (58/88) at phase
bin 0.05 and 61.9\% (39/63) at phase bin 0.45, compared to a maximum of 79.5\%
(58/73) at phase bin 0.25. The within-condition z-phase sensitivity range was 17.6
percentage points --- larger than the 13.2\,pp overall sensitivity reduction
attributable to the interval size itself. Bootstrap confidence intervals at the
phase extremes (0.05: [55.7--76.1]; 0.45: [54.4--71.9]) did not overlap with
those at the peak (0.25: [69.9--87.7]). These results are summarized in
Figure~\ref{fig:zphase}.

\subsection{Interpretation of the U-Shaped Z-Phase Profile}

The observed pattern at 5\,mm --- reduced sensitivity at both $\varphi \approx 0$
and $\varphi \approx 0.5$ relative to intermediate phase values --- needs comment.
The expected mechanism predicts maximal sensitivity loss at $\varphi = 0.5$, where
nodule signal is most equally distributed across two planes. The reduced sensitivity
at $\varphi \approx 0$ is not predicted by the simple partial-volume model and
likely reflects higher-order reconstruction effects. Specific candidates include
mismatch between the nominal slice interval and the actual slice sensitivity profile
width, nodules whose z-extent is smaller than the reconstruction interval and
therefore represented at the margin of a single plane rather than centered within
it, and discretization bias arising from imprecise z-origin alignment. Further
characterization of this bimodal pattern, stratified by nodule size and z-extent,
is warranted.

\subsection{Interval/Diameter Ratio Analysis}

When detection results were pooled across all three reconstruction interval
conditions and stratified by d/D, a clear sensitivity gradient emerged. Overall
sensitivity was 92.4\% for d/D $<$ 0.5 (n=633), 78.0\% for $0.5 \leq$ d/D $<$
1.0 (n=177), and 61.4\% for d/D $\geq$ 1.0 (n=114). Within the d/D $\geq$ 1.0
stratum, z-phase sensitivity varied from approximately 52\% at phase 0.45 to 65\%
at phase 0.05, consistent with the U-shaped profile observed at 5\,mm. In contrast,
the d/D $<$ 0.5 stratum showed a flat z-phase curve at $\sim$92\% with tight
confidence intervals.

\begin{figure}[ht]
\centering
\includegraphics[width=\textwidth]{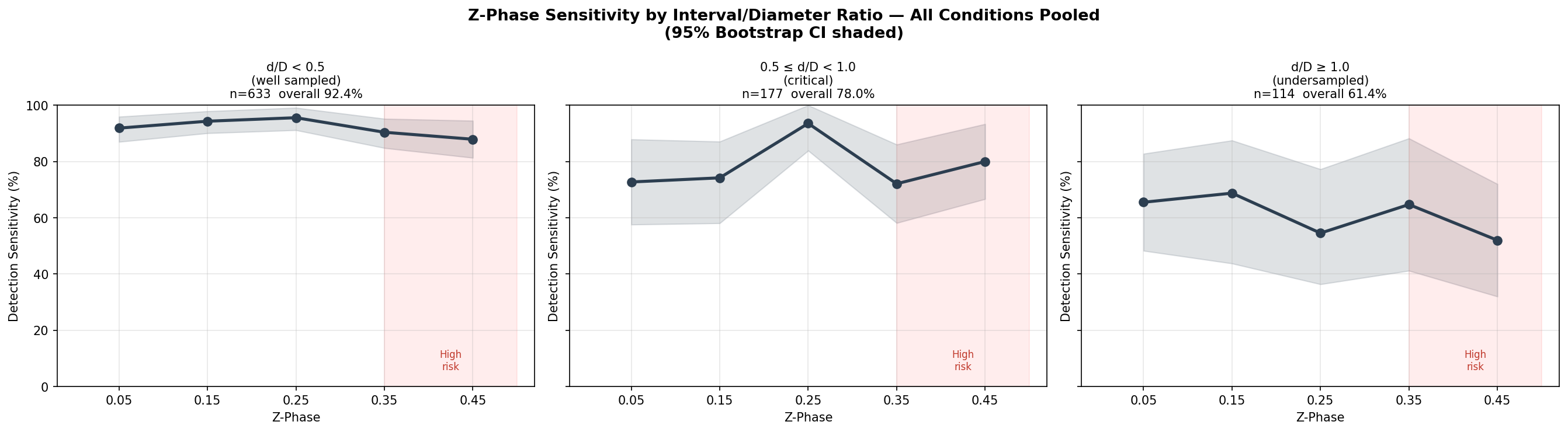}
\caption{\textbf{Interval/diameter ratio (d/D) analysis, pooling all reconstruction
interval conditions.} Each panel shows z-phase sensitivity with 95\% bootstrap
confidence intervals for a d/D stratum: well sampled (d/D $<$ 0.5, n=633),
critical range ($0.5 \leq$ d/D $<$ 1.0, n=177), and undersampled (d/D $\geq$ 1.0,
n=114). A systematic z-phase effect is present only in the undersampled stratum,
demonstrating that the positional effect is governed by the ratio of reconstruction
interval to nodule diameter rather than by interval size alone.}
\label{fig:ratio}
\end{figure}

These results indicate that the z-phase effect is governed by d/D rather than by
reconstruction interval size alone. The 5\,mm condition is the primary driver of
the d/D $\geq$ 1.0 stratum because coarser intervals more frequently exceed nodule
diameter in the 3--10\,mm clinical range. These results are summarized in
Figure~\ref{fig:ratio}.

\section{Discussion}

This study demonstrates that AI detection sensitivity for pulmonary nodules depends
not only on reconstruction interval size, but on the geometric relationship between
individual nodules and the reconstruction grid --- an effect we term \emph{z-phase
sensitivity dependence}. Although deterministic in principle, this effect is
effectively unpredictable at the study level because reconstruction grid offsets
and nodule anatomic positions are unknown at the time of AI inference. A secondary analysis introducing the
interval-to-diameter ratio d/D as the governing variable further characterizes the
conditions under which this effect is clinically meaningful.

\textbf{The effect is study-dependent and unpredictable from protocol alone.} The
reconstruction grid offset and the anatomic z-position of any given nodule are
effectively unpredictable in clinical deployment. Two studies acquired with
identical protocols can have z-phase values that differ by 0.25, corresponding to a
sensitivity difference of up to 17.6 percentage points in the 5\,mm condition. This
per-study variance is invisible to protocol-level quality dashboards.

\textbf{The effect is not reflected in AI confidence scores.} A nodule at z-phase
0.45 in a 5\,mm-reconstructed study is $\sim$18 percentage points less likely to
be detected than one at z-phase 0.25 --- but if detected, the confidence score is
uninformative about this geometric context.

\textbf{Clinical implications.} The ratio analysis identifies the conditions under
which z-phase sensitivity dependence is clinically meaningful: when d/D approaches
or exceeds 1.0. This threshold is reached for nodules in the 3--6\,mm range at
5\,mm reconstruction intervals --- precisely the Lung-RADS category 3 window where
a missed nodule results in no follow-up recommendation. Large nodules
($\geq$10\,mm, d/D $<$ 0.5 at 5\,mm) are largely unaffected, as are all nodule
sizes at 1\,mm or 3\,mm reconstruction. For studies where d/D $\geq$ 1.0, the
expected sensitivity is not a fixed 71.6\% but rather a distribution ranging from
approximately 62\% to 80\% depending on nodule position. For 3--6\,mm nodules at 5\,mm reconstruction,
detectability is not a fixed sensitivity but a distribution conditioned on where
the nodule happens to fall within the reconstruction cycle.

\textbf{Limitations.} The index-matching approach used to assign z-phase to
detection outcomes assumes annotation consensus ordering is consistent with the
original detection pipeline's nodule ordering, which may introduce minor
misalignment for cases with multiple nodules. The z-origin was derived from DICOM
\texttt{ImagePositionPatient} headers rather than from the NIfTI volumes used in
the original inference pipeline. Nodule diameter was estimated from XML annotation
bounding box extent rather than measured directly. The analysis is limited to a
single deep learning architecture (MONAI RetinaNet trained on LUNA16). The 3--6\,mm
size stratum at 3\,mm reconstruction (n=49) was underpowered to confirm or exclude
a z-phase effect at d/D $\approx$ 1.0.

\section{Conclusion}

AI detection sensitivity for pulmonary nodules varies systematically with the
geometric relationship between nodule position and the reconstruction grid. At
5\,mm reconstruction intervals, z-phase sensitivity varied by up to 17.6 percentage
points --- exceeding the mean sensitivity loss attributable to the interval size
itself. A ratio analysis pooling all conditions demonstrates that this effect is
governed by the interval-to-diameter ratio d/D: sensitivity was 92.4\% at d/D $<$
0.5 and 61.4\% at d/D $\geq$ 1.0, with a systematic positional effect present only
in the undersampled stratum. Scan-specific detectability characterization that
accounts for reconstruction interval z-phase and nodule size represents a previously
unrecognized dimension of AI reliability monitoring in CT lung nodule screening ---
one that is effectively stochastic at the study level and invisible to existing
protocol-level quality frameworks.

\section*{Acknowledgements}
Detection results analyzed in this study were generated using the MONAI framework
and LIDC-IDRI dataset. The author declares no conflicts of interest.


\end{document}